# Mechanism for Learning Object retrieval supporting adaptivity

Sonal Chawla and R.K. Singla

**Abstract** --In today's world designing adaptable course material requires new technical knowledge which involves a need for a uniform protocol that allows organizing resources with emphasis on quality and Learning. This can be achieved by bundling the resources in a known and prescribed fashion called Learning objects. Learning Objects are composed of two aspects namely "Learning " and "Object". The Learning aspect of Learning objects refers to Education. Since Education is a process so the primary aim of learning objects tends to be facilitating acquisition, assessment and conversion of content into Learning objects while fostering the assimilation of these Learning objects into learning modules and instruction. The Object part of Learning objects relates to the Digital Electronic format of the resources i.e. to say that it deals with the physical resource that forms the Learning objects. The objects in LOs are analogous to objects used in object-oriented modeling (OOM). The analogy helps visualize how LOs will be packaged, processed and transported across the digital library as well as utilized in course building. OOM concepts such as encapsulation, classification, polymorphism, inheritance and reuse can be borrowed to describe the operations on LOs in the digital library. Thus, the aim of this paper is threefolds. Firstly, to discuss the background of this research and the concept of Learning Objects. Secondly, to provide a framework for adaptive mechanism for the retrieval of Learning Objects and thirdly to highlight the benefits that this new proposed framework shall bring.

**KeyWords:** Learning Objects, Adaptive retrieval, Xerte, Moodle.

## 1  Introduction

Digital resources are widely used in today's education and their pace of growth is considerably fast. There has been consistent leaps forward in the methods to group, share, retrieve and reuse curricular information through software applications. These initiatives give support to emergent educational paradigm where concepts like learning, knowledge, skills, competencies, learning outcomes make-up the core of the current educational perspectives. [1] All these components can work independently and yet support learning and testing of its students. This can be achieved through the concept of Virtual University Learning Environment. [2]

But creating this environment involves the creation of effective E-learning resources which can be transferred for use on other platforms. This is called Interoperability. The challenge of making resources interoperable across different systems thus becomes a major task. [9]

The primary response to these problems is a major area of research with numerous international work directed at developing small reusable chunks of educational material in the form of Learning Objects. In this context, this study describes specific efforts to increase and diversify learner's access to E-learning applications. The strategies to offer the necessary physical and technological infrastructure to allow the designing and implementation of digital resources in the form of Learning Objects are discussed in this paper. The study also looks at pedagogical methodologies and research, evolvement and innovation in teaching and learning tools based on E-learning standards. More precisely, the study investigates specific ways in which digital resources are designed and stored and automatically delivered to learners.

This work is centered in methodological actions aimed to design, produce and distribute learning objects using instructional schemas and their performance in Virtual education environment built following instructional engineering approaches. The study identifies the components to design and produce these digital contents to be exploited in adaptive virtual environments. The applied research based on postulates and prototypes aims to consolidate the existent learning experiences using ICT to enhance the teaching and learning processes, promote the didactic innovation and add value to the research

────────────────────────────
- *Sonal Chawla is with the Dept of computer Science & Appl, Panjab University, Chandigarh. India.*
- *R.K. Singla is with the the Dept of computer Science & Appl, Panjab University, Chandigarh. India.*

initiatives. The transformation from traditional to digital curricula is accomplished considering the functional analysis model adapted to support academic program requirements. The Bloom's Cognitive Domain and the Felder learning style model premises are combined to allow in the curricular planning, the definition of instructional strategies and media needs that should guide the design and digital content production in the form of Learning Objects, focused to achieve meaningful and personalized learning in this emerging educational paradigm. [4]

The research provides an effective approach to the process of Learning Object Design and Construction. The Learning Object design and construction for this study was done using the Authoring software Xerte, launched recently by University of Nottingham and the dissemination of resources was done using Moodle. The Learning Objects were stored using MySQL database. The aim of this research then was to propose an adaptive mechanism to deliver the Learning Objects from the database suiting the learning style and the Learning preference of the learner. For this a framework has been proposed in the next section considering various learning styles and learning preferences of the learners. [6] The drivers for the introduction of Learning Objects is an identification to the changing needs of the students and resource redundancies involved with the traditional course development processes. This approach, however, is not about replacing the human dimension of teaching and learning experience, but about enhancing the face-to-face time spent between students and instructors. This research apparently enhances the quality of human interaction due to the fact that the time when students and teachers interact is not wasted by transmitting information which could be more effectively transmitted by other means. This research blends the four basic principles of Learning Objects: reusability, interoperability, durability and accessibility.

The reuse of digital learning material has been a continuing issue. First there were a number of initiatives promoting the reuse of educational software. However, their success in practice was limited. The most substantial problems were incompatibilities in language, culture, curriculum, computer-use practices, and pedagogical approaches of the potential learners and their instructors . Although David Wiley compared the idea of building educational content from smaller building blocks with object oriented programming [5] yet there is no generally agreed development and reuse concept in areas like software engineering. It has been argued by the researchers that design principles such as encapsulation, cohesion, and decoupling allow educators to develop and maintain objects independently of each other. Boyle was the first who attempted to transfer certain software engineering principles like cohesion and decoupling to Learning Objects to encourage the production of reusable Learning Objects. Cohesion among different components of a Learning Object in Boyle's [3] approach is achieved by the fact that all components are focused on a single learning objective.

## 2.0 Mechanism for adaptive delivery

However, the delivery of learning objects needs an Adaptive Learning System model that is based on constructivism and adaptive learning. Thus, there must be a mechanism that gathers information about the learners and matches it with digital resources from the Learning Object Repository thereby generating content, context and information that the learners need. This mechanism depicting the collaboration of digital resources with learner profiles has been shown in Figure 1
This part needs instructional designers and subject matter experts to collaborate using artificial intelligence principles, ontology, techniques, rules etc to develop a matching topology. Using this mechanism, adaptive or individualized learning objects can be retrieved from the repository following the aggregated data emerging from learner profiles, learning context and learning objectives. [8]
Adaptive Learning Objects have been discussed from various viewpoints with regards to learner's adaptation through interoperability issues. Recent researches on Learning Objects are contributing for the search of patterns for instructional content development, in order to make them adaptive, generic, portable and scalable enough to improve their potential for reusability.

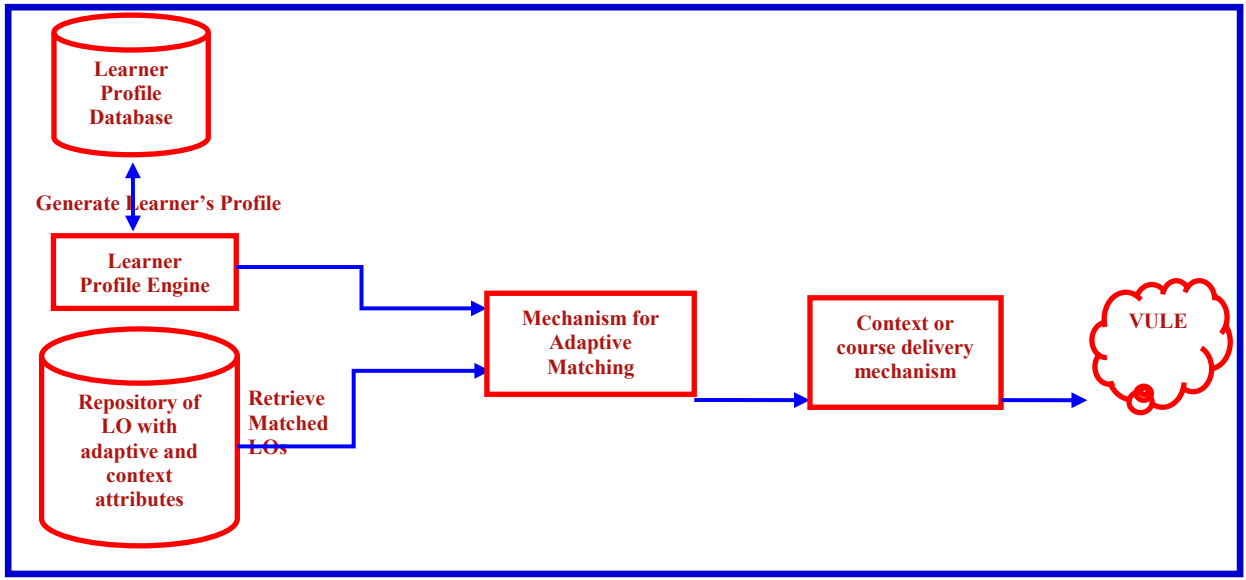

**Figure 1: Mechanism for Adaptive Matching of Learning Objects in VULE**

## 3.0 Architecture Proposal for Learning Objects supporting interactivity

Although after gathering the learning objects from the repository, they can be delivered by assembling into the course unit with context and presented to the learners using the Learning Management System yet the great challenge still remains as to how adaptive a learning object could be. The Figure 2 presents an architectural proposal for the learning objects supporting interoperability and adaptation.

This architecture for repositories proposed allows Learning Objects to be retrieved in an adaptive manner. This architecture is a six-layered architecture with the first tier as the Learner Profile tier. This tier holds responsibility for keeping track of student's historical information, each learner's individual profile repository, information about individual learning expectances and their learning goals. All these pieces of information are extremely important to provide adaptation in the historical, pre-requisite-based way. Several levels of adaptation can be established relating to the wide range of different aspects on teaching-learning processes. [10] These could range from keeping track of student's evolution on building some desired knowledge matching the learning styles picked up from the historical data as well as heuristic-based processes.

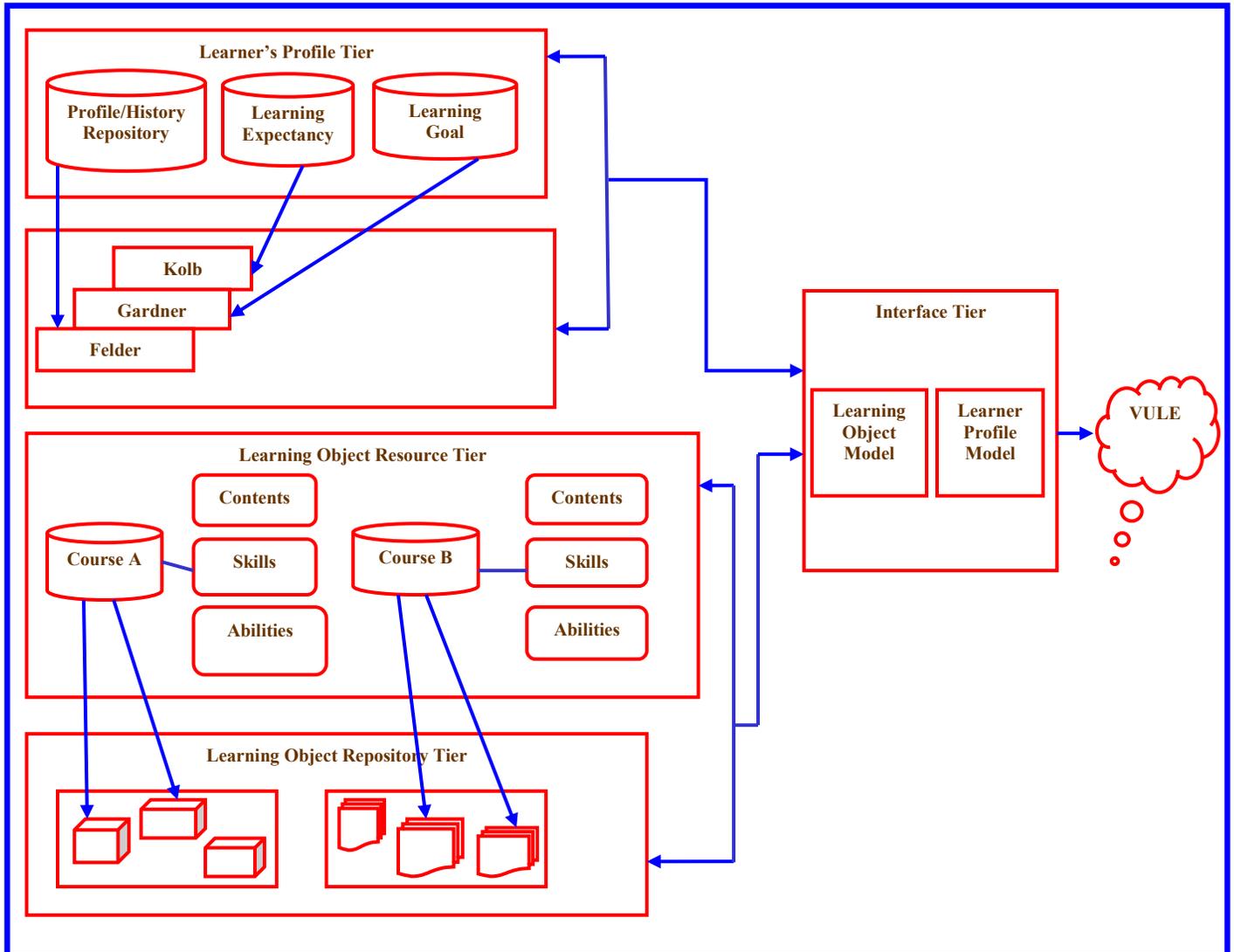

**Figure 2 : Six Layered Architecture for adaptive Learning Objects**

Second tier is the Learning Styles tier which sprawls in multi-dimensional measure and is influenced by a range of factors like environment, inter-relational issues and psychological aspects conveying how the student deals with knowledge to be constructed or skill to be developed. This work draws on research from Kolb, Felder and Gardner dwelling upon the classification of Learning styles. Student's profiles pass through the learning styles mentioned by these researchers to create the Learning style-dependent profile. It is pertinent to mention here that the Learning styles are not dependent only on student's historical profile but also of knowledge's nature. This makes them expected skills and abilities. This tier employs instructional design strategy for selection and sequencing of Learning Objects based on learning objectives. Since, here, learning styles based adaptation is to be considered so adaptive heuristics must decide which objects are to be shown to certain students in some learning context. Fourth tier is the Learning Object repository tier. Here the Learning Objects are organized in the repositories. They could be custom produced or retrieved from other interoperable repositories. But the gathered Learning Objects must be interoperable relying on the three techniques of Interoperability i.e. federated searches, LO harvesting through metadata and finally LO gathering. This becomes possible when the repositories, from where the data is being retrieved, follows standards. These standards are extremely essential for categorization and classification of Learning Objects in their discovery and recovery process.

Fifth tier is the Interface tier. This tier dynamically generates the suitable visualization of the Learning Objects according to the Learner's profile based on the learning styles for a certain learning context. This tier is composed of Learning Object Module and the Learner Profile Module. The Learning Object Module describes all the Learning Objects to be used in the learning context based on basic sequencing by Instructional Design. The Learner Profile Module contains the learner's historical profile and the learning styles. Both these modules match the Learning Objects as per the learner's profile using mechanism to display them through the Virtual University Learning Environment.

Sixth tier is the Virtual University Learning Environment for disseminating the Learning Objects.

## 4.0  Results and Discussions

This new learning environment is expected to facilitate:
- Creation of learning objects during course to facilitate personalised learning
- Use of advance organisers to enhance assimilation
- Automated formative assessment to scaffold learning
- Constructivist social learning to promote teamwork
- Inquiry-based and exploratory learning to synthesize knowledge

The Model aims to provide a Virtual University design that connects its learners in an innovative learning environment, providing them with access to a wide range of formal and informal community groups, including discussion forums enabling students to interact, collaborate and support each other, sharing their individual experiences as they progress through their studies. Besides others, it offers benefits like:-

- Faculty across departments can independently create and produce online content, while maintaining the uniform look and feel of Virtual Campus.
- Improved departmental and college portals enhance student and faculty recruiting, alumni relations and development efforts.
- New workflow process and organization-wide shared content trims costs and boosts employee productivity by minimizing the time and effort needed to create and manage online content.
- Better online content, improved organization and consistent design helps ensure students have quick and easy access to the information they need.

http://www.ijklo.org/Volume1/v1p229-254Kay_Knaack.pdf